# ONE-SIDED MUON TOMOGRAPHY – A PORTABLE METHOD FOR IMAGING CRITICAL INFRASTRUCTURE WITH A SINGLE MUON DETECTOR

K. Boniface[1], V.N.P. Anghel, A. Erlandson*, G. Jonkmans[2], M. Thompson, S. Livingstone

Canadian Nuclear Laboratories Ltd, Chalk River, Ontario, Canada

## Abstract

High-energy muons generated from cosmic-ray particle showers have been shown to exhibit properties ideal for imaging the interior of large structures. This paper explores the possibility of using a single portable muon detector in conjunction with image reconstruction methods used in nuclear medicine to reconstruct a 3D image of the interior of critical infrastructure such as the Zero Energy Deuterium (ZED-2) research reactor at Canadian Nuclear Laboratories' Chalk River site. The ZED-2 reactor core and muon detector arrangement are modeled in GEANT4 and Monte Carlo measurements of the resultant muon throughput and angular distribution at several angles of rotation around the reactor are generated. Statistical analysis is then performed on these measurements based on the well-defined flux and angular distribution of muons expected near the surface of the earth. The results of this analysis are shown to produce reconstructed images of the spatial distribution of nuclear fuel within the core for multiple fuel configurations. This one-sided tomography concept is a possible candidate for examining the internal structure of larger critical facilities, for example the Fukushima Daiichi power plant where the integrity of the containment infrastructure and the location of the reactor fuel is unknown.

## 1.     Introduction

The non-destructive examination of the internal composition of an object is often desirable – whether it be a human body, a cargo container at a sea port or something as large as a reactor core. In the case of the human body, advancements in nuclear medicine have made it possible to reconstruct anatomical images by actively administering small doses of radiation such as X-rays to a patient and measuring its distribution as it exits the body [1]. Larger structures, on the other hand, generally have a much higher probability of completely stopping this type of radiation, thus requiring a different methodology.

High-energy cosmic-ray muons have been used for non-destructive imaging for decades, with imaging methods ranging from 2D analysis using measurements of muon attenuation [2] to 3D 'muon tomography' based on the multiple Coulomb scattering (MCS) of muons as they pass through material [3,11]. To produce a 3D image, the muon tomography method requires two detectors to measure the muon trajectory before and after transiting the structure, thus allowing the scattering angle of individual muons through the structure to be determined. Analysis of the magnitude and spatial distribution of muon scattering events can provide a means of imaging the internal composition of structures with particular emphasis on discriminating between materials of high, medium and low atomic number [3][11].

---
[1] Current address: McMaster University, Hamilton, Ontario, Canada
[2] Current address: Defence R&D Canada – Centre for Security Science, Ottawa, Ontario, Canada
* Corresponding author: andrew.erlandson@cnl.ca

This paper aims to demonstrate the effectiveness of a muon imaging technique that requires only one detector for examining critical infrastructure for the purpose of nuclear safeguards, waste management, and assessment of damaged reactor cores. Using a single portable detector that measures the distribution of the attenuated muon flux through a structure from several points of view, a 3D image of the structure's internal composition can be reconstructed, with particular emphasis on the presence and spatial distribution of high-Z material. The overall concept is analogous to the type of imaging performed in nuclear medicine, with particle detection occurring only after its transmission through the object in question, and the reconstruction of an image based solely on its measured projections [1]. As such, our ongoing research investigates the integration of the algebraic reconstruction technique (ART) – an iterative reconstruction method commonly used in medical imaging [1] – with the detection and analysis of muons after their passage through a structure. Unlike medical imaging that has a known controlled radiation source, this proposed technique relies on background muon radiation that has a range of directions and energies.

In Section 2, we begin by comparing the proposed 'one-sided muon tomography' concept to the established two-sided scattering tomography technique [11]. The basic mathematical concepts of tomography in the field of medical imaging are then reviewed in Section 3, and it is shown that the proposed concept is closer to the original meaning of tomography than two-sided muon tomography. Section 4 then presents reconstructed images based on Monte Carlo measurements of muons after their passage through a simple mock-up of the ZED-2 research reactor with different configurations of nuclear fuel within the core. In particular, the fuel distributions examined are arranged to simulate a reactor in a post-accident scenario. The reconstructed distribution of nuclear fuel throughout the core is clearly distinguishable in each case, supporting the validity of this imaging technique for further application.

## 2. Imaging with Muons

Muons are charged particles created as a result of the interaction of cosmic radiation – mainly high-energy protons – with the upper layer of the earth's atmosphere. They can be detected near the surface of the earth at a rate of approximately 1 cm$^{-2}$min$^{-1}$ with an average energy of 3-4 GeV [4]. The high energy of muons coupled with their large mass ($\approx 206\ m_e$) [4] enables them to penetrate deeply into matter before being attenuated, hence they provide the means to image large structures containing high-density, high-Z material [3]. As muons traverse through matter, they lose energy due to ionization and experience MCS causing small changes in their trajectory [4], and each of these physical processes provide the basis for different imaging techniques.

### 2.1 Multiple Coulomb Scattering and Two-Sided Muon Tomography

The total scattering angle of a muon through any material is directly related to the atomic number and density of the material and the momentum of the incident muon. Statistically, the distribution of the planar scattering angle of muons through a material of a given thickness is approximately Gaussian with a mean of zero and a standard deviation that increases as the density and atomic number of the material is increased and as the momentum of the muon is decreased [4]. Due to the high energy and large mass of muons, the scattering that occurs is relatively small in magnitude (i.e., mrads), enabling individual muon trajectories to be readily tracked by placing detectors directly before and after the object in question [5][11].

Figure 1 shows a schematic of the Cosmic Ray Inspection and Passive Tomography (CRIPT) prototype system which is currently operating at Chalk River Laboratories (CRL) to exploit the MCS of muons for the purpose of imaging nuclear fuel for safeguards and waste management [5][11]. From the directional information recorded for each detected muon, a 3D representation of the density of scattering locations throughout the cargo bay area can be produced and the nature of the scattering material can then be inferred. This concept allows for the resolution of low, medium and high-Z material located within the cargo, and is particularly sensitive to special nuclear material (SNM) such as uranium and plutonium [5][11].

## 2.2 Attenuation and One-Sided Muon Tomography

In the field of nuclear medicine, X-ray computed tomography (CT) is a type of diagnostic scan that involves irradiation with X-rays to generate images of a patient's anatomy. An X-ray source and detector arrangement are rotated around the central axis of the patient, taking measurements of the distribution of transmitted X-rays on the exit side of the body. At each discrete angle of rotation, the measured data represents a 2D projection of the 3D distribution of X-ray attenuation coefficients – or tissue types – within the patient. The objective is then to reconstruct the 3D distribution based solely on the measured projections [1].

Measuring the attenuation of muons as they pass through matter can be used in a similar manner if the scale of the object is large enough. As muons pass through a given material, ionization causes them to lose energy at a rate of approximately 2 MeV per $g/cm^2$ [4]. Consequently, with initial energies in the GeV range, significant attenuation of the muon flux will be observed only after the particles have traversed large quantities of high-density material such as that found in a reactor core.

The imaging system used during a CT scan requires a detector on only one side of the patient; a second detector measuring incoming radiation is not required as a known source is used. The measured distributions are interpreted by comparing them to the uniform distribution measured if the patient were not present [1]. In order to be truly analogous to an X-ray CT scan, a muon imaging system requires only one detector that detects muons after their transmission through a structure, and the detector must be moved to various locations around the central axis of the structure. This concept requires knowledge of the distribution of muons expected without the structure present.

The angular distribution of muons near the surface of the earth has been well-documented, and can be described as having uniformity in the azimuthal angle, and a zenith angle (deviation from vertical) distribution that is dependent on the energy of the particles. For the average muon with an energy of ~3 GeV, the zenith angle distribution is $cos^2\theta$ in shape, where $\theta$ is the angle between the muon trajectory and the vertical after projection onto either of the two vertical orthogonal planes [4]. Using this information, it becomes possible to eliminate the necessity of an 'upper detector' (referring to Figure 1) in a muon imaging system.

Using a horizontal detector (such as that shown in Figure 1) it would be difficult to find a suitable axis of rotation around many large structures, hence the initial proof of concept of the one-sided muon tomography method uses a vertical detector that can be positioned to the side and below a structure. The non-vertical components of the muon flux at several angles of rotation around the structure can then be detected and analyzed. Figure 2 illustrates an example of the geometrical setup for this imaging concept, with the structure being a large vertical cylinder, representing a simplification of the ZED-2 reactor core. At each angle of detection around the central axis of the core, the expected flux in each angular component of the distribution is compared with the measured flux, providing a measurement of the relative attenuation in each direction that can yield information about the type of material the muons passed through.

## 3.  2D Image Reconstruction From Projections – Mathematical Background

### 3.1  Forward Projection and the Radon Transform

The Radon transform is an integral transform that provides the mathematical foundation for tomographic image reconstruction. The transform describes the process of projection by relating a 2D function to the collection of line integrals of that function along paths normal to the plane of projection [6]. Figure 3 illustrates an example of this concept with a simple density function $f(x,y)$ shown in the $(x,y)$-plane and its projection $R_\theta(x')$ onto the $x'$-line that has been rotated at an angle $\theta$ with respect to the positive $x$-axis.

At each position on the $x'$-line, the value of $R_\theta(x')$ represents the integral of $f(x,y)$ along a line parallel to the $y'$-axis and passing through $x'$. Thus the Radon transform of a function can be described mathematically by [6]:

$$R_\theta(x') = \int_{-\infty}^{\infty} f(x,y) dy' \qquad (1)$$

In the case of X-ray CT, the function $f(x,y)$ maps the X-ray attenuation coefficient at any point $(x,y)$ in the cross-section of the patient and the function $R_\theta(x')$ is the number of X-ray events detected at a position $x'$ with respect to the rotated coordinate system of the detector [1]. If $R_\theta(x')$ is known for an infinite number of values of $\theta$, inversion of the Radon transform can be used to recover the original density function from its projections [1][6]. In medical imaging, it is common practice to reconstruct individual 2D cross-sections of a patient independently and combine these slices to produce a final 3D image [1].

In reality $R_\theta(x')$ and $f(x,y)$ are functions of discrete variables – there are finite points of measurement (called detector bins) along the $x'$-axis and finite image pixel positions in the $(x,y)$-plane. As such, a system of algebraic equations can be constructed that describes the array of unknown pixel densities in the cross section of the image in terms of the projection data from all angles of measurement [6]. For each individual bin on the detector plane and for each angle of detection, the weighted contribution of every pixel in the image plane to that measurement can be modeled. With a total of $N$ detector measurements and $M$ pixels on the image plane, the system of equations is described by [6]:

$$r_i = \sum_{j=1}^{M} A_{ij} f_j \quad where\ i = 1, 2, \cdots N \tag{2}$$

where the vector $\boldsymbol{r}$ holds the entire dataset of detector measurements, the vector $\boldsymbol{f}$ is a one-index representation of the pixels on the image plane, and the matrix $\boldsymbol{A}$ is a discretization of the Radon transform with each element $A_{ij}$ representing the weight that is assigned to pixel $j$ for the $i^{th}$ element in the measured dataset.

### 3.2    The Algebraic Reconstruction Technique (ART)

While inversion algorithms do exist for the Radon transform of continuous functions, the discrete nature of image reconstruction creates significant challenges that preclude the use of these algorithms [7]. Hence, a way of approximating the inversion process must be used instead. This paper focuses on a particular iterative reconstruction method (ART) which assigns an initial estimate to each pixel and updates the solution based on a comparison of the projections calculated from the current image estimate to the measured projection [1].

This method is of row-action type whereby each of the equations that describe the system is examined individually; the vector of unknowns is updated after each equation has been analyzed and then the next equation is addressed. One iteration of the algorithm is complete when every equation in the system has been addressed once. The process can be described mathematically by the following formula (based on the information presented in [1]):

$$\boldsymbol{f}^{(k+\frac{i}{N})} = \boldsymbol{f}^{(k+\frac{i-1}{N})} + \frac{r_i - \langle A_{(i,\cdot)}, \boldsymbol{f}^{(k+\frac{i-1}{N})} \rangle}{\|A_{(i,\cdot)}\|^2} A_{(i,\cdot)}^T \tag{3}$$

where $k$ is the iteration number, $A_{(i,\cdot)}$ is the $i^{th}$ row of the matrix $\boldsymbol{A}$ and $i$ goes from 1 to $N$, allowing for each equation in the system to be analyzed in sequential order for each iteration.

The process is initiated by assigning an estimate $\boldsymbol{f}^{(0)}$ to the solution vector – this is typically a uniform vector with all elements having a value of zero [1]. At each step, the projection value that corresponds to the current solution estimate $r_{i,current\ estimate} = \langle A_{(i,\cdot)}, \boldsymbol{f}^{(k+\frac{i-1}{N})} \rangle$ is calculated and compared to the actual measured projection value, $r_i$.

The result of this comparison is then used to modify the current estimate of the solution $f^{(k+\frac{i-1}{N})}$ to create a new estimate $f^{(k+\frac{i}{N})}$. The process is then repeated until a specified convergence condition has been met [1].

## 4. Simulations of the ZED-2 Research Reactor

The ZED-2 research reactor located at AECL CRL is an ideal candidate for testing muon imaging methods. It is a small, zero-power, heavy-water moderated reactor in which a broad range of fuel types and fuel channel configurations can be studied [8]. The next two sub-sections discuss results from the previously studied two-sided muon tomography method and present preliminary results from the proposed one-sided approach.

### 4.1 Previously Published Results Using Two-Sided Tomography

In 2013, Jonkmans *et al.* published the results of a feasibility study in which the two-sided muon tomography method (as discussed in section 2.1) was theoretically demonstrated for the purpose of imaging critical facilities using a simplified simulation of the ZED-2 geometry [9]. Two large (9x4 $m^2$) vertical detectors were placed on either side of the reactor, as shown in Figure 4 (left) and two different fuel channel configurations (Figure 4, right) were simulated and reconstructed.

By analyzing the scattering angles of individual muon tracks over a period of 5 days, the images shown in Figure 5 were reconstructed [9]. The difference in fuel mass distribution is clearly discernible for the two cases, thereby supporting the validity of the concept; however it is important to consider the size and placement of the detectors in regards to the feasibility of the experiment. When the space constraints inside the reactor building are taken into account, it is clear that these 9x4 $m^2$ detectors cannot realistically be positioned as suggested by Figure 5. Interestingly, this two-sided method is currently being proposed for imaging the Fukushima Dai-ichi power plant to assess the damage done by the earthquake in 2011 [10].

### 4.2 Preliminary Results for the Proposed One-Sided Tomography Method

Using instead the proposed one-sided muon tomography method, a smaller, portable detector can be utilized and provide a more practical option for imaging the contents of critical infrastructure such as the ZED-2 core. The detector is 1x1 $m^2$ and can therefore be manoeuvred more easily within the confined spaces of the reactor building. Figure 6 shows one possible configuration of detector positions where the detector (shown in red) is placed in the space between the graphite reflector and concrete wall and can be rotated around the central axis of the reactor. In comparison, the arrangement shown in Figure 4 is not physically possible due to constraints from equipment and infrastructure.

There are two significant geometrical considerations to be made when attempting to use the image reconstruction methods described in Section 3 in conjunction with measurements of cosmic-ray muons. First, given the natural distribution of muons at the surface of the earth, it is difficult to separate the imaging volume into 2D cross-sections to be reconstructed independently, and so a full 3D imaging volume must be considered from the start. Second, we are no longer dealing with projection rays that are all parallel at each angle of detection.

In the reconstruction methods used in medical imaging, each bin on the detector expects to see particles with only one incident angle [1], [6]. A muon detector however observes particles with a range of incident angles. Since the distribution of muons at any point near the surface of the earth is expected to be the same over a long enough period of time, a solution to this problem is to divide each bin on the plane of detection into further sub-bins corresponding to the angular components of the distribution seen by the detector. An illustration of this angular bin concept is shown in Figure 7 where the detector (shown in blue) has been divided into a number of bins (called pixels), and for each pixel, the detected muons are then divided further into bins corresponding to their incoming angle. Each individual angular bin (an example of which is highlighted in red in Figure 7) corresponds to a separate equation in the system described by Equation 2. Consequently the total number of equations, $N$, is equal to:

$$N = nD \times nP \times nA \qquad (4)$$

where $nD$ is the number of detector positions used, $nP$ is the number of pixels the detector is divided into, and $nA$ is the number of angular bins the observed muon distribution has been divided into.

Another notable difference between using X-rays and muons for imaging is the amount of scattering experienced by each type of particle. In a CT scan, the majority of the X-rays either pass through the tissue are absorbed; the amount of scattering that occurs is small by comparison [1]. With muons, the relative amount of scattering versus attenuation is more significant [4] and is consequently a source of noise in the reconstructed image. However, similar to the imaging method described in Section 2.1, the one-sided tomography method relies on the fact that the magnitude of the scattering angles is relatively small due to the high energy of the particles [4]. The two-sided method makes use of this by only requiring a detector system with a small angular acceptance [5]. The one-sided method assumes that since these deflections are small that the corresponding level of noise will also be small. In the next section, we will see that this is a valid assumption given that the fuel mass distributions are easily distinguishable in each presented case.

### 4.2.1 One-Sided Tomography Results

Figure 8 (top row) shows four different fuel configurations within the ZED-2 core and the resulting reconstructed images (middle row) obtained using the one-sided muon tomography method. Twelve angles of detection spaced evenly over 360° were used (refer to Figure 6 for positioning) and at each angle, approximately 2.75 days worth of muon events were simulated (resulting in a total exposure time of ~33 days). The dataset obtained at each position of the detector was then subdivided with values of $nP = 100$ and $nA = 358$ resulting in a total of 429,600 equations in the system, each corresponding to an individual element in the vector $r$ (as described in Section 3.1). For each fuel configuration, the data recorded in $r$ is the number of muon events detected in each angular bin subtracted from the number of events that were expected in each bin without the reactor geometry present. In this way, the measured flux in each angular component is normalized to the expected, non-uniform $cos^2\theta$ distribution found in nature.

After a single iteration of the ART algorithm (Equation 3), the images shown in the middle row of Figure 8 were reconstructed. In each of the top-down projections of the reconstructed images, the overall distribution of nuclear fuel is clearly visible. The outer dimension of the calandria vessel (which contains heavy water) is observed to be close to its true value of a radius of 168 cm and the relative location of the fuel within the vessel is in agreement with the simulated channel locations.

In each of the images, two inaccuracies can be noted. The first is the low-density that has been reconstructed at the centre of the core. The second is the oscillating pattern of high-density material seen where the fuel is present. It is of note that in the reconstructed images of Figure 8, there are 12 iterations of this oscillating pattern spaced evenly over 360°. It is for this reason that they are believed to correspond to the 12 angles of detection that have been used. Both inaccuracies are believed to be an artefact of the weighting matrix used during the ART process and a method for correcting this is currently being investigated.

A method of background subtraction was implemented in an effort to highlight the effectiveness of the one-sided tomography method in light of the inaccuracies previously mentioned. The reconstructed image of the normal operations fuel configuration (Figure 8, left column) was used as a control case. Each of the subsequent reconstructed images were subtracted from this control case resulting in the images shown in the bottom row of Figure 8. This method neutralizes the effects of the low-density centre and the oscillations within the reconstructed fuel and as a result, clearly highlights the placement of missing fuel within the core. However, this requires an image of the structure before any 'damage' occurs; this initial image may not be available and therefore efforts are underway to improve the image reconstruction algorithms to remove these artefacts.

## 5. Conclusion

The preliminary results presented in this paper show that using a single, portable muon detector can provide a more practical method to image the internal composition of critical infrastructure than the two-sided tomography method. Given careful consideration to the manner in which the raw detector measurements are processed, one-sided muon tomography is able to distinguish between materials of high, medium and low atomic number and in particular, can indicate when SNM is not present in places it was assumed to be. However, further development of the image reconstruction technique is required to eliminate the requirement for creating a 'before' image of the critical infrastructure.

Two-sided muon tomography has been suggested as a candidate for imaging large critical infrastructure such as operating reactor cores [9], nuclear waste storage containers [3], and for assessing the damage done to the Fukushima Dai-ichi power plant in 2011 [10]. However, the size and cost of the required detectors is significantly greater than that of the proposed one-sided method. While the total detection time for the one-sided approach is longer than two sided (~ 33 days versus ~ 5 days), the overall practicality of manipulating and placing the smaller portable detector(s) (and subsequently expected lower costs) of the one-sided method suggests that it may be the more feasible option. In certain applications where the limits on detection time are not strict (e.g., imaging large, stationary structures for the purpose of nuclear waste management and non-proliferation), the proposed one-sided tomography approach may be a more practical option.

Development of the reconstruction techniques described in this paper continues. In particular, focus will now be given to optimizing parameters such as the size, placement and orientation of the detector, the size of the binning parameters (detector pixels and angular bins), correcting the weighting matrix, and refining the convergence criteria during the iterative solving process. The simulation of a more realistic detector, and the use of real data to further verify the validity of the concept will also be pursued.

## 6. Acknowledgements

The authors would like to acknowledge the productive cooperation between McMaster University and Canadian Nuclear Laboratories which facilitated the student internships which contributed to this work.

## 8. Figures

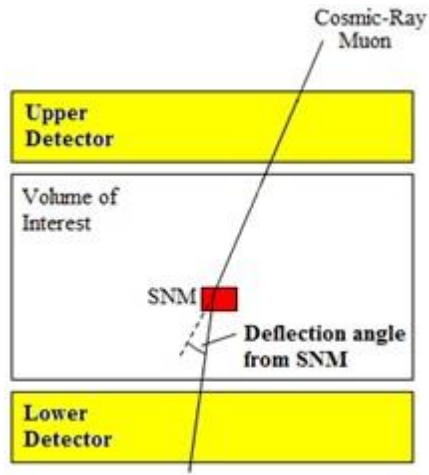

**Figure 1:** A schematic of the CRIPT imaging system which utilizes the two-sided muon tomography concept.

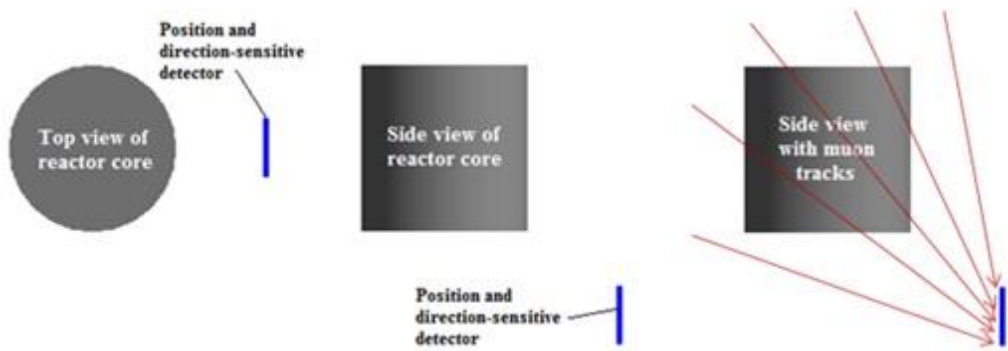

**Figure 2:** Geometrical setup for the one-sided muon tomography concept.

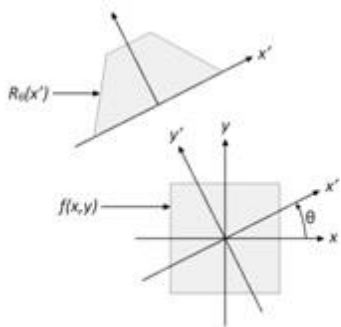

**Figure 3:** Projection of a simple density function onto a rotated coordinate system.

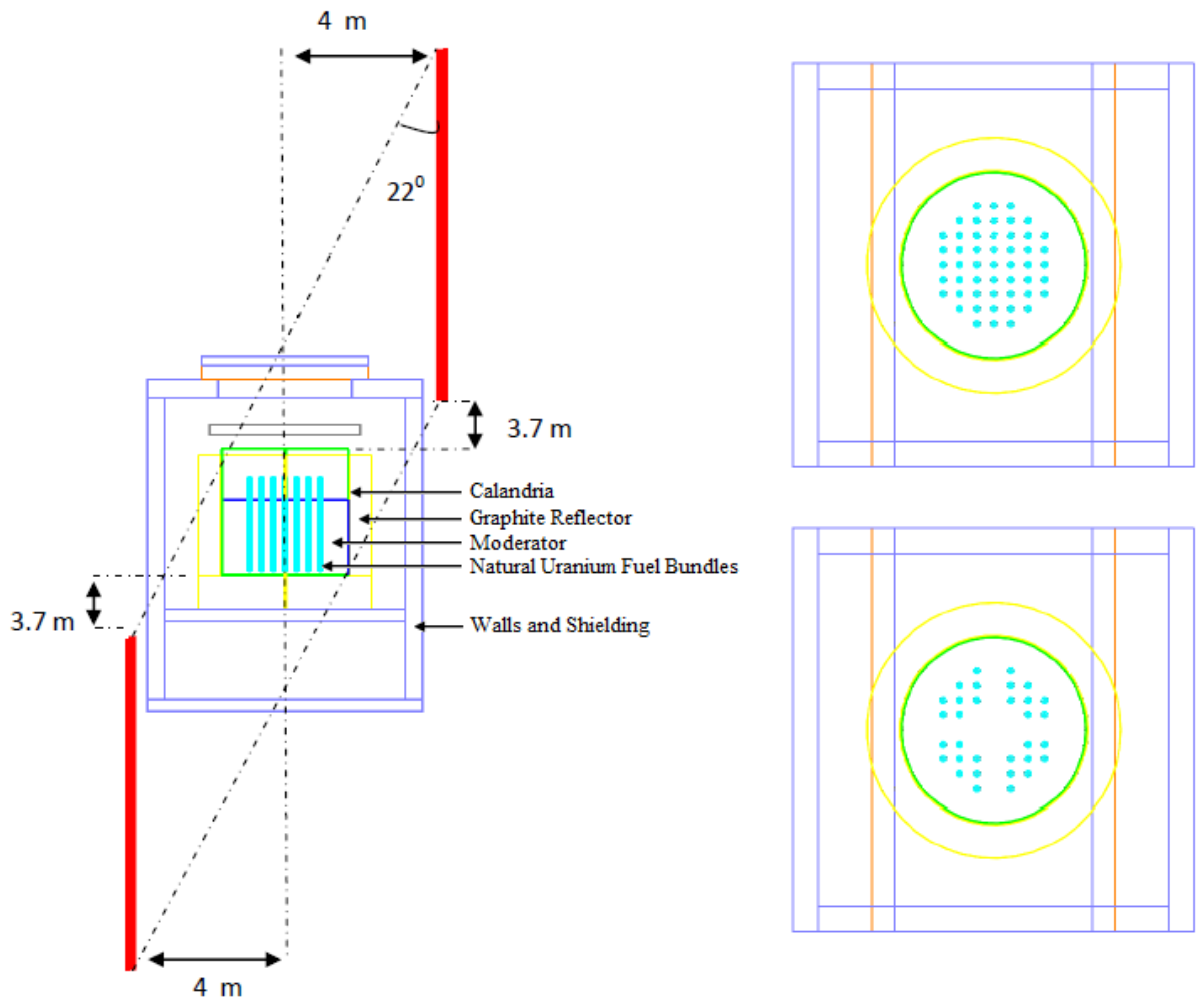

**Figure 4:** Left: main structural components of the ZED-2 reactor with two vertical muon detectors (thick red lines) placed on either side of the core. Top right: top-view of a fuel configuration representing normal operating conditions. Bottom right: top-view of a fuel configuration representing an accident scenario in which 19 fuel channels are missing [9].

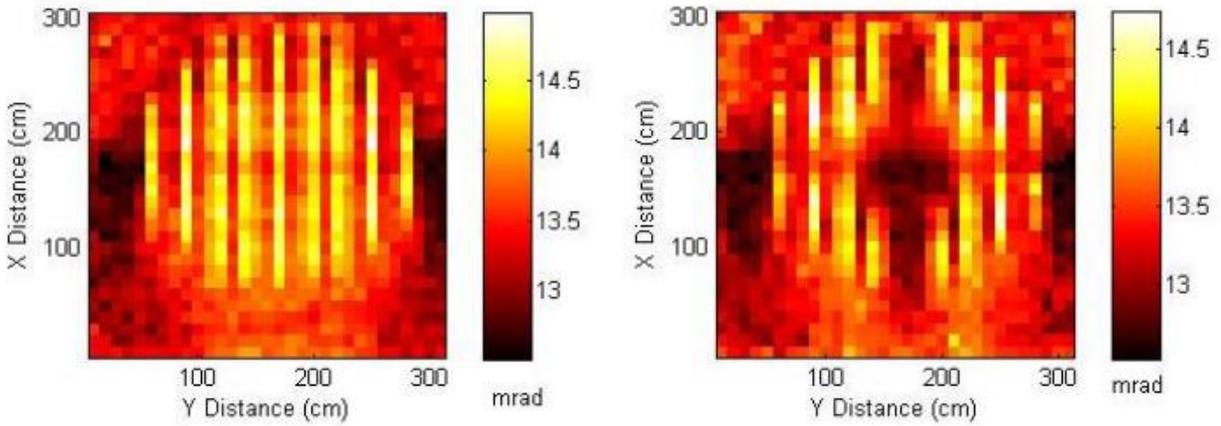

**Figure 5:** Reconstructed images of ZED-2 core during normal operating conditions (left) and a postulated accident scenario (right) [9]. Both images were reconstructed using two-sided muon tomography (refer to Figure 4).

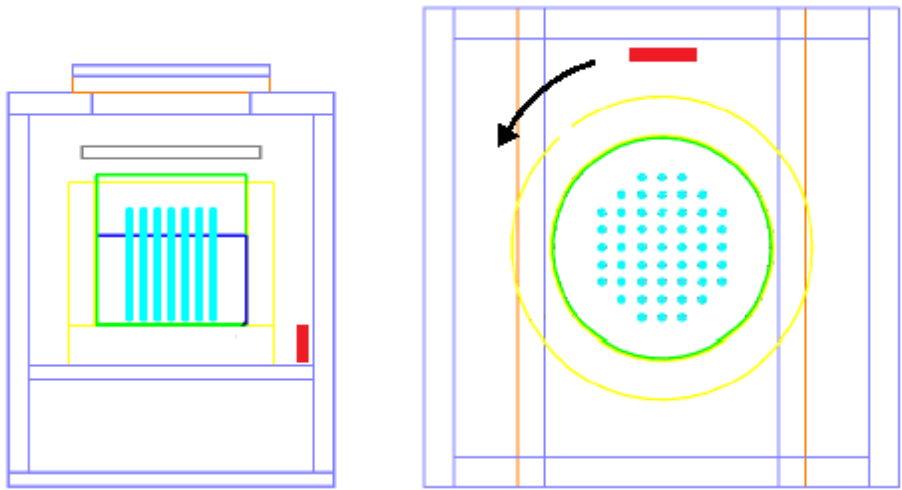

**Figure 6:** Detector placement for one-sided muon tomography of the ZED-2 reactor. Left: side-view. Right: top-view.

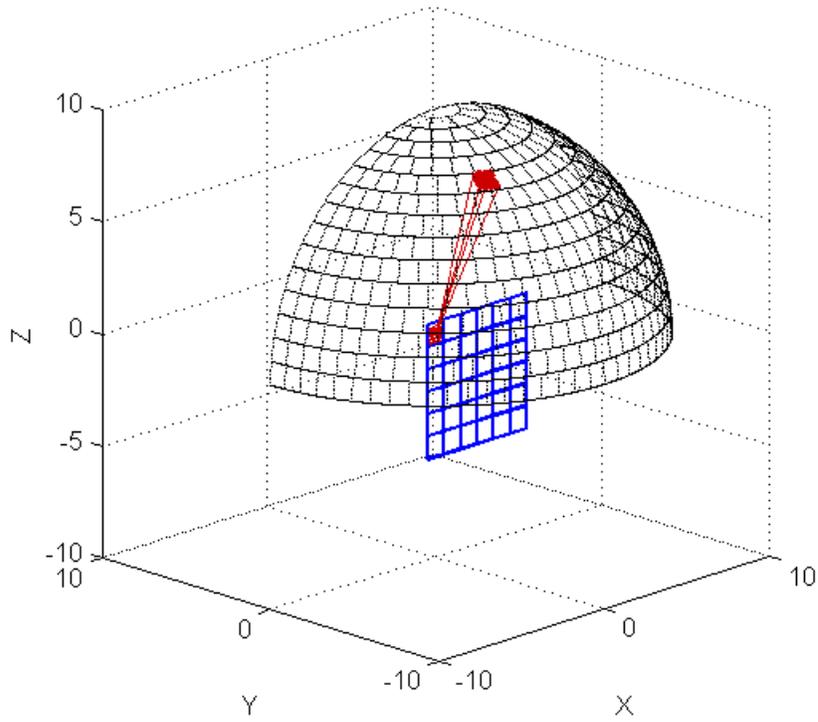

**Figure 7:** Illustration of the division of muon events into angular bins. Vertical muon detector shown in blue. A single angular bin has been highlighted in red.

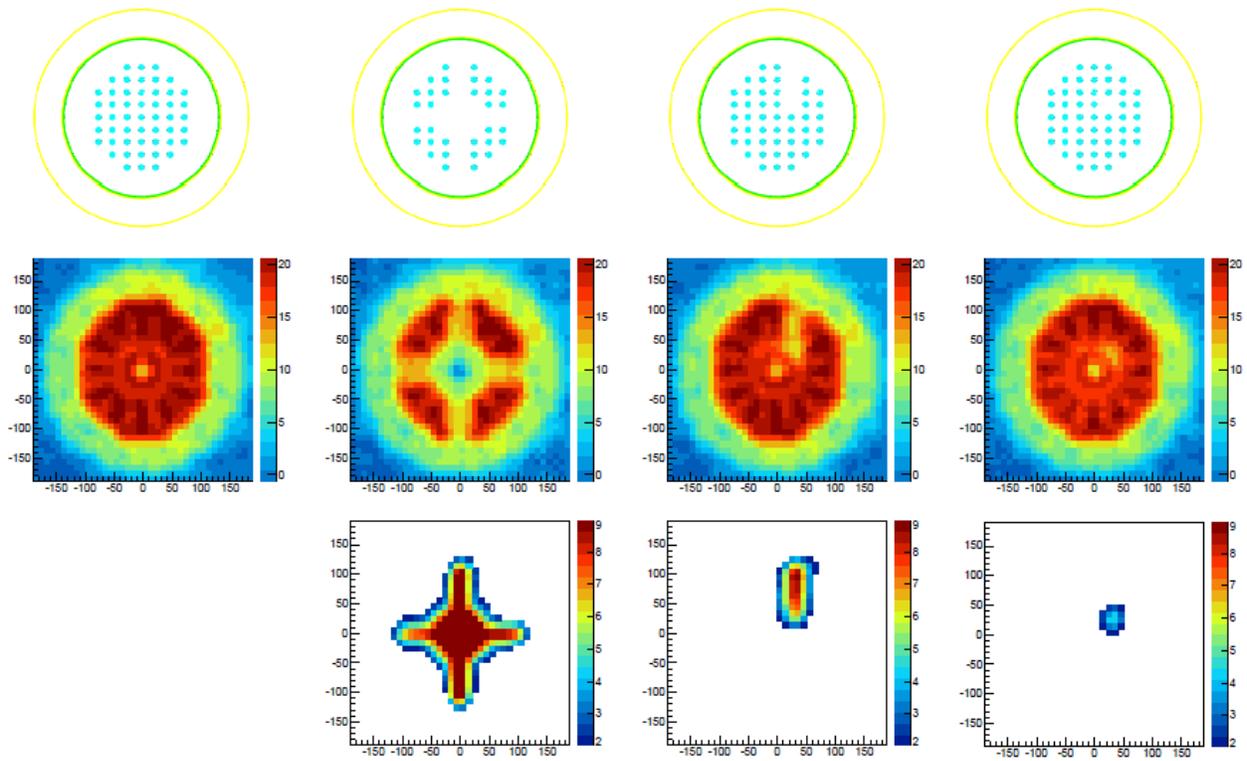

**Figure 8:** Top row: top-down projection of four different fuel channel configurations. Middle row: reconstructed images using 12 angles of detection and 1 iteration of the ART algorithm. Bottom row: background-subtracted images in which the leftmost fuel configuration was used as the control case; the missing fuel is clearly visible in each case.